\documentclass[11pt,a4paper]{article}

\usepackage{epsfig,amsmath,amssymb,cite}

\begin{document}

\title{Circularly symmetric thin-shell wormholes in F(R) gravity \\ with (2+1)-dimensions}

\author{Cecilia Bejarano\thanks{e-mail: cbejarano@iafe.uba.ar}, Ernesto F. Eiroa\thanks{e-mail: eiroa@iafe.uba.ar}, Griselda Figueroa-Aguirre\thanks{e-mail: gfigueroa@iafe.uba.ar}\\
{\small  Instituto de Astronom\'{\i}a y F\'{\i}sica del Espacio (IAFE, CONICET-UBA),}\\
{\small Casilla de Correo 67, Sucursal 28, 1428, Buenos Aires, Argentina}} 
\date{}
\maketitle

\begin{abstract}

Within the framework of $F(R)$ theories of gravity with (2+1)-dimensions and constant scalar curvature $R$, we construct a family of thin-shell wormholes with circular symmetry and we analyze the stability of the static configurations under radial perturbations. We show an example of asymptotically anti-de Sitter thin-shell wormholes with charge, finding that stable configurations with normal matter are possible for a suitable range of the parameters.

\end{abstract}

\textit{Keywords: Wormholes, F(R) gravity, Thin shells.}

\section{Introduction}\label{sec: intro}

One of the main goals of alternative theories of gravity is focused on some fundamental issues in the current cosmological picture, from the early-time inflation scenario to the late-time cosmic accelerated expansion. Basically, the intention is to modify the geometrical side of the gravitational field equations without adding any unknown matter-energy content. Certainly, any alternative approach also intends to provide a geometrical explanation for the spacetime singularities as well as a cornerstone for  quantum version of gravity theory.   There is a plethora of different modified gravity models. In particular, $F(R)$ gravity \cite{sot10,fel10,noj17} is one of the most straightforward modifications to General Relativity (GR) since the Einstein-Hilbert action is replaced by
\begin{equation}
S=\frac{1}{2\kappa} \int d^{4}x \sqrt{|g|} \left(F(R)+ \mathcal{L}_m \right)  ,
\end{equation}
with $\kappa =8\pi$, and $R$ is the scalar curvature. We adopt units such that c = G = 1, with c the speed of light and G the gravitational constant. In the metric formalism, in which the metric tensor is the only dynamical variable to perform the variation of  the action, the field equations read
\begin{equation}
F'(R)R_{\mu\nu}-\dfrac{g_{\mu\nu}}{2}F(R)-\nabla_\mu\nabla_\nu F'(R)+g_{\mu\nu}\nabla_\rho\nabla^{\rho} F'(R)=\kappa T_{\mu\nu} ,
\label{eq:fieldeq}
\end{equation}
where $\nabla$ is the covariant derivative and $T_{\mu \nu}$ is the energy-momentum tensor.

Here, we present charged thin-shell wormholes in $F(R)$ gravity in (2+1) dimensions, finding stable static solutions for certain values of the parameters. The construction follows the junction formalism \cite{isr66a, isr66b, der08, sen13} which is also applied, for instance, to study thin layers of matter around (i) vacuum (namely, bubbles) or (ii) black holes. Bubbles made of pure double layer in quadratic $F(R)$ have been studied a few years ago \cite{gfa3}; and layers of matter surrounding black holes have also been explored \cite{gfa51,gfa52,gfa53}. The stability of (3+1)-dimensional thin-shell wormholes in $F(R)$ gravity has been recently investigated \cite{gfa1,gfa2}.

\section{Junction conditions}\label{sec: jc}

The cut-and-paste formalism used to construct some wormhole geometries is based on the junction conditions of the theory by allowing to match different solutions across a hypersurface $\Sigma$, which represents a boundary if there is not matter-energy content or a thin shell of matter otherwise. We adopt the notation that the jump of any quantity $\Upsilon $ across $\Sigma$  is defined by $[\Upsilon ]\equiv (\Upsilon ^{2}-\Upsilon ^{1})|_\Sigma$. 

\subsection{General Relativity}

In GR, the junction conditions are expressed by the Darmois-Israel formalism \cite{isr66a,isr66b} which can  be sketched as follows:
\begin{itemize}
\item 
Two manifolds  $\mathcal{M}_1$ and $\mathcal{M}_2$ are glued at a hypersurface $\Sigma$ to give  ${\cal M}=\mathcal{M}_1 \cup \mathcal{M}_2$.
\item 
The first fundamental form  (the induced metric) at $\Sigma$ is
\begin{equation}
h^{1,2}_{ij}= \left. g^{1,2}_{\mu\nu}\frac{\partial X^{\mu}_{1,2}}{\partial\xi^{i}}\frac{\partial X^{\nu}_{1,2}}{\partial\xi^{j}}\right| _{\Sigma} ,
\end{equation}
where  ${X^\mu _{1,2}}$ are the coordinates that describe $\mathcal{M}_{1,2}$, respectively; and  ${\xi ^a}$ are the intrinsic local coordinates at the hypersurface, hence  $X^\mu _{1,2}=X^\mu _{1,2} (\xi^a)$. 
\item 
The second fundamental form (the extrinsic curvature) at $\Sigma$ is given by
\begin{equation}
 K_{ij}^{1,2}=-n_{\gamma }^{1,2 }\left. \left( \frac{\partial ^{2}X^{\gamma}_{1,2} } {\partial \xi ^{i}\partial \xi ^{j}}+\Gamma _{\alpha \beta }^{\gamma }
\frac{ \partial X^{\alpha }_{1,2}}{\partial \xi ^{i}}\frac{\partial X^{\beta }_{1,2}}{\partial \xi ^{j}}\right) \right| _{\Sigma } , 
\end{equation}
with  $n_{\gamma }^{1,2 }$ the unit normal ($n^{\gamma }n_{\gamma }=1$) pointing from $\mathcal{M}_{1}$ to $\mathcal{M}_{2}$.
\item The matching between $\mathcal M_1$ and $\mathcal M_2$ at $\Sigma$, should fulfill the so-called junction conditions:
\subitem -- 
The line element should be continuous across $\Sigma$: 
\begin{equation}
[h_{\mu\nu}] = 0 . 
\end{equation}
\subitem -- The field equations at $\Sigma $ (Lanczos equations) reduce to  
\begin{equation}\label{lanczoseq}
\kappa S_{\mu \nu}=-[K_{\mu \nu}]-h_{\mu \nu}K^\gamma_{\;\;\gamma} ,  \quad n^{\mu}S_{\mu\nu}=0
\end{equation}
\subitem where $S_{\mu \nu}$ is the energy-momentum tensor at the gluing hypersurface.
\end{itemize}

The hypersurface $\Sigma$ is called a \textit{boundary surface} if $S_{\mu \nu} = 0$,
or \textit{thin shell} if $S_{\mu \nu} \neq 0$.

\subsection{$F(R)$ gravity}

The formalism for the junction conditions in $F(R)$ gravity was developed in recent years \cite{der08,sen13}. In this theory, there are two separate cases to take into account.

\paragraph{General (non-quadratic) case:} $F'''(R) \neq 0$
\\

At the matching hypersurface $\Sigma$, the continuity of the following quantities should be accomplished
\begin{equation}
[h_{\mu\nu}]=0, \quad [K^\mu{}_\mu]=0, \quad [R]=0 , 
\end{equation}
and the field equations result
\begin{equation}
\kappa S_{\mu\nu}=-F'(R)[K_{\mu\nu}]+F''(R)[n^{\rho}\nabla_{\rho}R]h_{\mu\nu} , \quad n^{\mu}S_{\mu\nu}=0 .
\label{eqfieldnonquad}
\end{equation}

\paragraph {Quadratic case:} $F'''(R) = 0$ 
\\

In this case, there is one less restriction and the  discontinuity of $R$ is allowed
\begin{equation}
[h_{\mu\nu}]=0, \quad [K^\mu{}_\mu]=0 .
\end{equation}
By taking $F(R)=\alpha R^2+R-2\Lambda$, the field equations read
\begin{equation}
\kappa S_{\mu \nu} =-[K_{\mu\nu}]+2\alpha( [n^{\gamma }\nabla_{\gamma}R] h_{\mu\nu}-[RK_{\mu\nu}]), \quad n^{\mu}S_{\mu\nu}=0 .
\label{eqfieldquad}
\end{equation}
In order to  guarantee the divergence-free energy-momentum tensor (locally conservation) at $\Sigma$, it is necessary to contemplate the following extra contributions \cite{sen14,rei16}

\noindent $-$ \textit{External energy flux vector}: 
\[ \kappa\mathcal{T}_\mu=-2\alpha\nabla_\mu[R], \quad n^{\mu}\mathcal{T}_\mu=0 , \] 
\noindent $-$ \textit{External scalar pressure/tension}: 
\[ \kappa\mathcal{T}=2\alpha [R] K^\gamma{}_\gamma , \]
\noindent $-$ \textit{Double layer contribution}: 
\[ \kappa \mathcal{T}_{\mu \nu}=\nabla_{\gamma } \left( 2\alpha [R] h_{\mu \nu } n^{\gamma } \delta ^{\Sigma }\right) , \] 
with $ \delta ^{\Sigma } $ the Dirac delta on $\Sigma$ (which is analogous to the dipole distributions in classical electrodynamics). It is easy to verify that these contributions vanish if the scalar curvature is continuous at $\Sigma$.

\section{Charged solutions with constant curvature}\label{sec: cc}

$F(R)$ gravity coupled to a nonlinear electromagnetic field  is described by the action\cite{hen14}
\begin{equation}
S=\frac{1}{2 \kappa }\int d^3x \sqrt{|g|} (F(R)+ (-\mathcal{F}_{\mu\nu}\mathcal{F}^{\mu\nu})^s) ,
\label{eq:action}
\end{equation}
where $s$ is a positive parameter ($s\neq 1/2$) and  $\mathcal{F}_{\mu \nu }=\partial _{\mu }\mathcal{A}_{\nu } -\partial _{\nu }\mathcal{A}_{\mu }$ is the electromagnetic tensor. The corresponding energy-momentum tensor becomes traceless for $s = 3/4$, so this value is adopted in what follows. 

The dynamical equations given by Eq.~(\ref{eq:fieldeq}) for the gravitational sector are completed with the electromagnetic field part
\begin{equation}
\partial_\mu \left(\sqrt{-g}F^{\mu\nu}(-F_{\alpha\beta}F^{\alpha\beta})^{-1/4}\right)=0.
\end{equation}
For constant $R=R_0$, the spherically symmetric geometry \cite{hen14}
\begin{equation}
ds^2=-A (r) dt^2+A (r)^{-1} dr^2+r^2d\theta^2 , 
\label{eq:sphsym}
\end{equation}
with 
\begin{equation}
A(r)=-M-\frac{\left( 2\mathcal{Q}^{2}\right)^{3/4}}{2 F'(R_0) r}-\frac{r^{2}R_{0}}{6} ,
\label{eq:A(r)}  
\end{equation}
is the solution of the theory represented by the action (\ref{eq:action}), where $R_{0}=6 \Lambda_{eff}$ can be thought in terms of an effective cosmological constant, $M$ is the mass and $Q$ is the charge. If $\mathcal{Q}=0$, the vacuum static BTZ geometry of GR is obtained. Therefore, this is also a solution in $F(R)$ gravity. Notice that this spacetime is asymptotically anti-de Sitter for $R_0<0$. 

For future purpose, we define an effective charge with its sign determined by the sign of $F'(R_0)$
\begin{equation}
Z \equiv \frac{\left( 2\mathcal{Q}^{2}\right)^{3/4}}{2 F'(R_0)} .
\end{equation}
Different values of $Z$ establish the horizon structure of the geometry given by (\ref{eq:A(r)}).  From this spacetime, in the next Section, we construct thin-shell wormholes by applying the junction formalism in $F(R)$ gravity and we study the stability by following the  standard potential approach \cite{gfa4}. 

\section{Wormhole geometry: construction and stability}\label{sec: whs}

The thin-shell wormhole spacetime is built  by glueing the two side of the spherically symmetric solution (\ref{eq:sphsym}) on $\Sigma$, which is defined by $G(r_{1,2})=r-a=0$, to create a geodesically complete manifold $\mathcal{M}=\mathcal{M}_{1}\cup \mathcal{M}_{2}$ where each side is given by $\mathcal{M}_{1,2}=\{ X^{\alpha}=(t,r,\theta) / r \geqslant a \}$: 
\begin{itemize}

\item{The area $4\pi r^2$ is minimal when $r=a$, so $\mathcal{M}$ represents a wormhole connecting two regions with a throat of radius $a$ (flare-out condition). }

\item $\xi ^{i}=(\tau ,\theta )$ are the coordinates on $\Sigma$, being $\tau $ the proper time. 

\item The thin-shell radius is taken as a function of time $a=a(\tau)$, in order to analyze the stability of the static shells.
\end{itemize}
Then, the metrics at the sides of the throat are given by
\begin{equation}
ds_{1,2}^2=-A_{1,2} (r) dt^2+A_{1,2} (r)^{-1} dr^2+r^2d\theta^2 , 
\end{equation}
with
\begin{equation}
A_i (r) =  1-\frac{2M_i}{r}+\frac{Q_i^2}{ F'(R_i) r^2}-\frac{R_i r^2}{12} , \quad i={1,2}
\end{equation}

In the orthonormal basis $\{ e_{\hat{\tau}}=e_{\tau }, e_{\hat{\theta}}=a^{-1}e_{\theta }\}$,
at the hypersurface $\Sigma$, the first fundamental form results
\begin{equation}
h^{1,2}_{\hat{\imath}\hat{\jmath}}= \mathrm{diag}(-1,1) ,
\end{equation}
and the non-null components of the second fundamental form  for static configurations are
\begin{equation}
K_{\hat{\tau}\hat{\tau}}^{1,2} = \pm \frac{A '_{1,2}(a_0)}{2\sqrt{A_{1,2}(a_0)}} , \quad
K_{\hat{\theta}\hat{\theta}}^{1,2} = \mp \frac{1}{a_0}\sqrt{A_{1,2} (a_0)} ,
\end{equation}
where the prime represents derivative with respect to $r$. The condition $[K^{\hat{\imath}}_{\;\; \hat{\imath}}]=0$, gives the following relation 
\begin{equation} 
\frac{A_{2}'(a_0)}{2\sqrt{A_{2}(a_0)}}+\frac{ A_{1}'(a_0)}{2\sqrt{A_{1}(a_0)}} + \frac{1}{a_0}\left(\sqrt{A_{1}(a_0)}+\sqrt{A_{2}(a_0)}\right) = 0 ,
\label{condstat}
\end{equation}
that should be always fulfilled by the throat radius $a_0$. 

We can obtain the dynamics of the thin shell from the time dependent version of $[K^{\hat{\imath}}_{\;\; \hat{\imath}}]=0$. Then, we analyze the stability of the configuration by rewriting the dynamical equation of the throat in terms of the potential 
\begin{equation}
\dot{a}^{2}=-V(a) ,
\end{equation}
where
\begin{equation}
V(a)= -\frac{a^2 \left(A_1(a)-A_2(a)\right)^2}{4 a_0^2 \left(\sqrt{A_1\left(a_0\right)}+\sqrt{A_2\left(a_0\right)}\right)^2} 
 -  \frac{a_0^2 \left(\sqrt{A_1\left(a_0\right)}+\sqrt{A_2\left(a_0\right)}\right)^2}{4 a^2} + \frac{A_1(a)+A_2(a)}{2} . 
\end{equation}
It is straightforward to see that $V(a_0)=0$ and  $V'(a_0)=0$.  Then, the static solutions are stable under radial perturbations when $V''(a_0)>0$, otherwise they are unstable. 

\begin{figure}
\includegraphics[width=1.\textwidth]{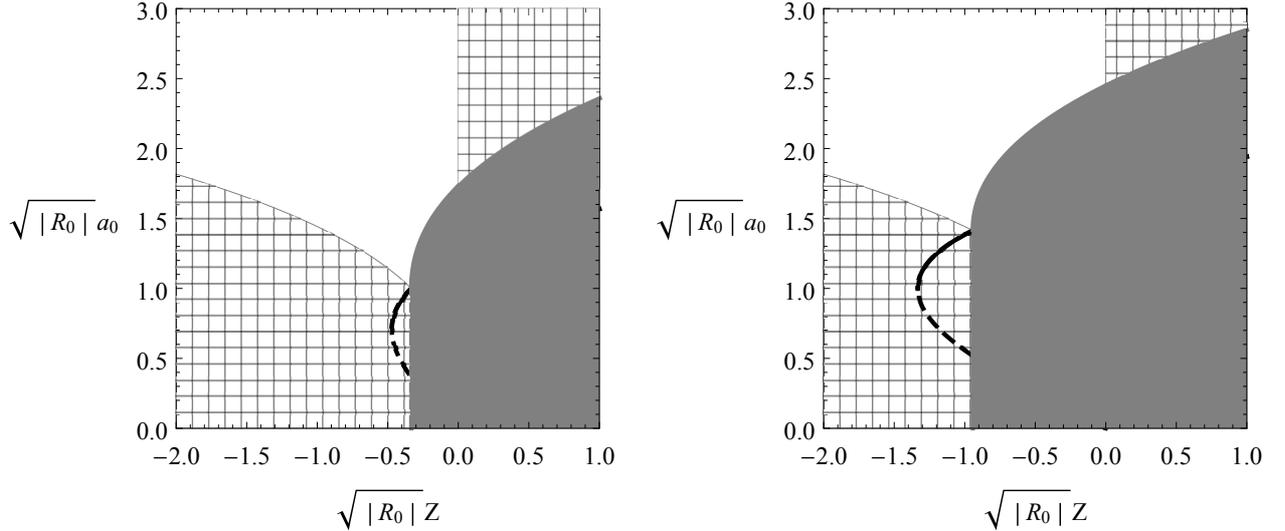}
\caption{Wormholes in general $F(R)$ gravity symmetric across the throat. Solid lines represent stable solutions while dashed ones unstable solutions. Meshed zones represent matter satisfying WEC. Grey zones have no physical meaning.  Left plot: $M=0.5$; right one: $M=1$.}
\label{fig1}
\end{figure}

\begin{figure}
\centering
\includegraphics[width=1.\textwidth]{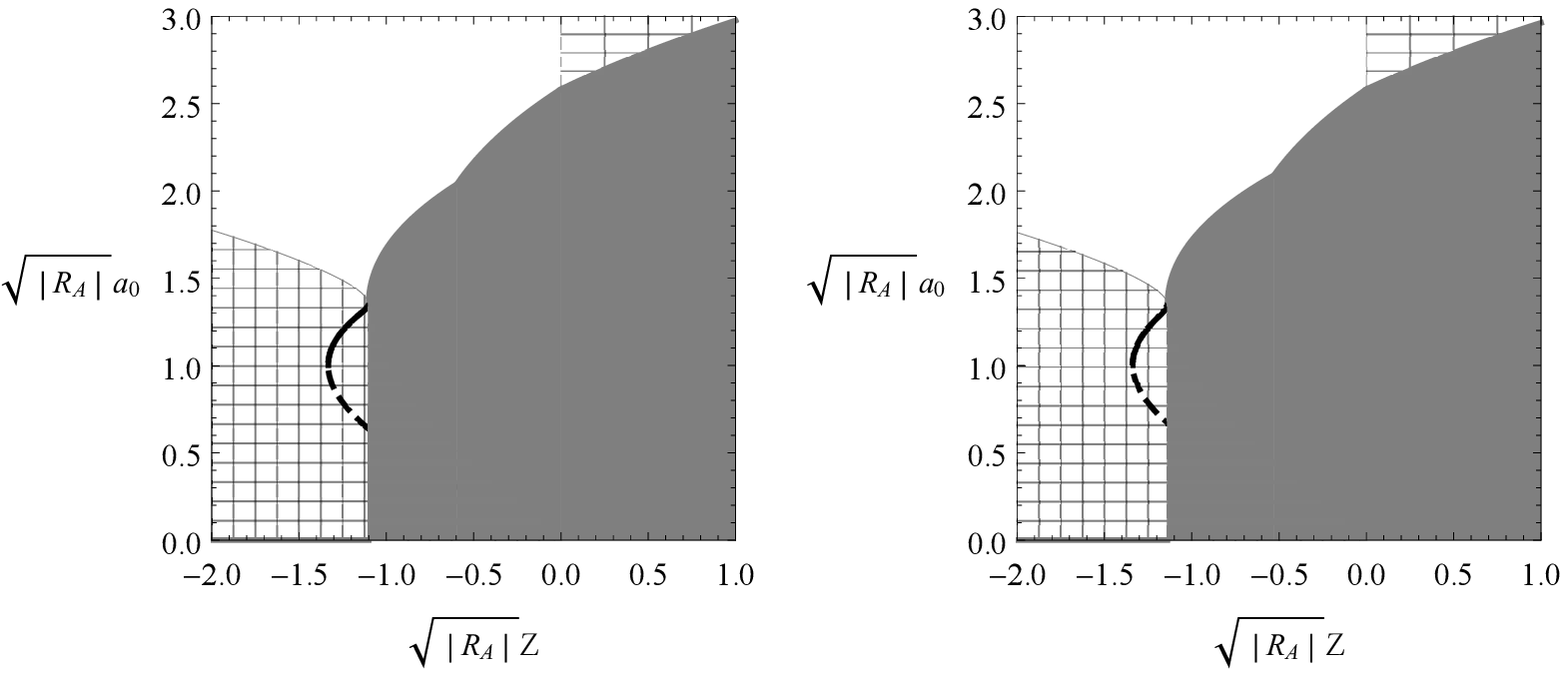}\\
\vspace{0.2cm}
\includegraphics[width=1.\textwidth]{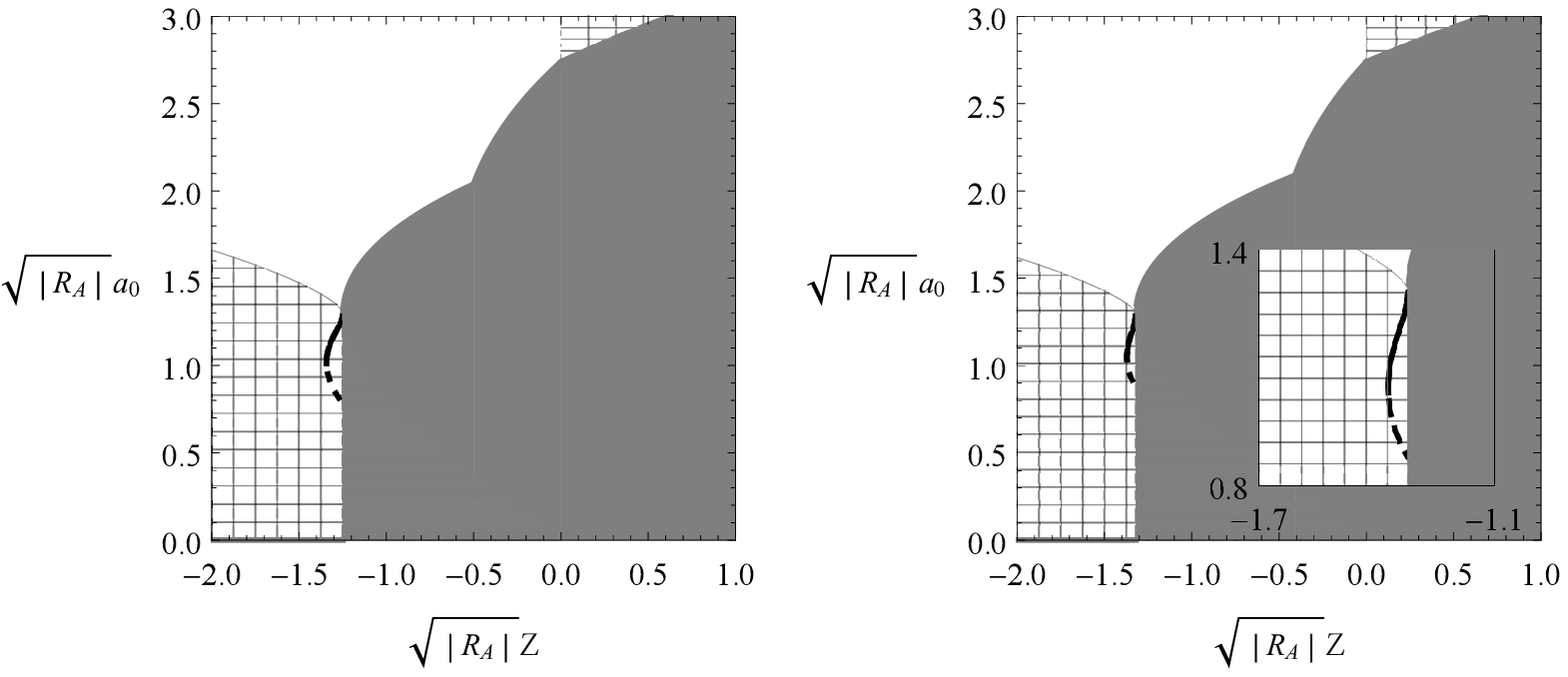}
\caption{Wormholes in quadratic $F(R)$ gravity with $R_1\neq R_2$ at the sides of the throat. Solid lines represent stable solutions while dashed ones unstable solutions. Meshed zones represent matter satisfying WEC. Grey zones have no physical meaning. For all plots: $M=1$, $R_A=(R_1+R_2)/2$. Upper row: $R_1 =0.9 R_A$ and $R_2 =1.1 R_A$; lower row: $R_1 =0.8 R_A$ and $R_2 =1.2 R_A$. Left column: $\alpha R_A=-0.1$ for $Z>0$ and $\alpha R_A =-0.9$ for $Z<0$; right column: $\alpha R_A=-0.2$ for $Z>0$ and $\alpha R_A =-0.8$ for $Z<0$.}
\label{fig23}
\end{figure}

\subsection{Wormholes in general $F(R)$ with $[R]=0$}

In the first example, we adopt a symmetric construction across the throat, i.e. $R_1=R_2=R_0$, $M_1=M_2=M$, and $Q_1=Q_2=Q$ so that $A_1(r)=A_2(r)=A(r)$. Then, from Eq. (\ref{eqfieldnonquad}), by using that $S_{\hat{i}\hat{j}}=diag(\sigma_0, p_0)$, we get the surface energy density and the pressure for the static solutions   

\begin{equation} 
\sigma_0 = \frac{F'(R_0)}{\kappa}\frac{A'(a_0)}{\sqrt{A(a_0)}} ,  
\end{equation}
\begin{equation}
p_0 =-2\frac{F'(R_0)}{a_0\kappa}\sqrt{A(a_0)} .
\end{equation}
If  $\sigma_0 <0$ or $\sigma_0 +p_0 <0$, the matter at the throat is exotic, because it does not satisfy the weak energy condition (WEC). The continuity of the trace of the extrinsic curvature imposes that the matter should satisfy the equation of state $\sigma_0 -2p_0=0$.

From Fig.~\ref{fig1}, we see that there are two solutions constituted of normal matter (satisfying the energy conditions) for a short range of $\sqrt{|R_0|} Z$ with $Z<0$. Hence it requires the presence of ghost fields, i.e. $F'(R_0)<0$. The solution with the largest radius is stable while the other one is unstable under radial perturbations. The qualitative behavior of the solutions is not affected by different values of $M$ which only represent a change of scale.

\subsection{Wormholes in quadratic $F(R)$ with $[R] \neq 0$}

In the second example, we adopt  $M_1=M_2=M$, and $Q_1=Q_2=Q$ but $R_1\neq R_2$, from Eq. (\ref{eqfieldquad}) the energy density and pressure at the throat in the static configuration are  
\begin{equation} 
\sigma_0 = \frac{1+2\alpha R_2}{\kappa }\left( \frac{A_{2}'(a_0)}{2\sqrt{A_{2}(a_0)}}\right) +  \frac{1+2\alpha R_1}{\kappa }\left( \frac{A_{1}'(a_0)}{2\sqrt{A_{1}(a_0)}} \right) \,  
\end{equation}
\begin{equation}
p_0  =-\frac{1+2\alpha R_2}{\kappa }\left( \frac{\sqrt{A_{2}(a_0)}}{a_0}\right) - \frac{1+2\alpha R_1}{\kappa }\left( \frac{\sqrt{A_{1}(a_0)}}{a_0}\right) .
\end{equation}
In this quadratic case, there exists an external scalar pressure/tension $\mathcal{T}_0$ that satisfies the equation of state $\sigma_0 -p_0=\mathcal{T}_0 $ which results from the condition $[K^\mu{}_\mu]=0$. The other extra contributions are: $\mathcal{T}_\mu^{(0)}=0$ and $\mathcal{T}_{\mu\nu}^{(0)}  \propto (2 \alpha [R]/\kappa) h_{\mu \nu}$.

From Fig.~\ref{fig23}, we see that there are two solutions made of normal matter for a short range of $\sqrt{|R_A|} Z$ with $R_A=(R_1+R_2)/2$. Since $Z<0$, there are ghost fields. The solution with the largest radius is stable; the other is unstable. The larger the difference between $R_1$ and $R_2$, the smaller the range of $\sqrt{|R_A|}Z$ where solutions exist. The same qualitative behaviour is obtained when  $\alpha$ decreases.

\section{Summary}

We have constructed a family of circularly symmetric wormholes in (2+1)-dimensional $F(R)$ gravity with constant scalar curvature $R$ at the sides of the throat. We have studied their stability under perturbations preserving the symmetry. The junction conditions in $F(R)$ gravity are more restrictive than in GR. In fact, the extra condition $[K^\mu{}_\mu]=0 $ forces the type of equation of state: for  general $F(R)$, $\sigma_0 -2p_0=0$, and for quadratic $F(R)$,  $\sigma_0 -2p_0=\mathcal{T}_0$. We have analyzed two examples: (i) symmetric wormholes ($R_1=R_2=R_0$; $M_1=M_2$; $Q_1=Q_2$)  across the throat in general $F(R)$ and (ii) wormholes asymmetric in the scalar curvature ($R_1\neq R_2$; $M_1=M_2$ and $Q_1=Q_2$) only in quadratic $F(R)$. In both cases we have found two solutions made of normal matter (i.e., satisfying WEC) for a range of values with $Z<0$. However, for these solution the presence of ghost fields is always required since $Z<0$ implies $F'(R_0)<0$.  We found that the solution with the larger radius is stable while the other is unstable. In particular, for quadratic $F(R)$ as long as the difference between the constant (negative) scalar curvatures $R_1$ and $R_2$ grows, the range of the squared charge $\mathcal{Q}^2$ where stable solutions exist becomes smaller; similar behavior can be observed when $\alpha $ decreases.

\section*{Acknowledgments}

This work has been supported by CONICET and Universidad de Buenos Aires. C. B. thanks the partial support of the John Templeton Foundation.

\end{document}